\documentclass[prl,twocolumn,superscriptaddress,showpacs,letterpaper]{revtex4}
\usepackage{graphicx}
\usepackage{natbib}
\def\EtMiss {E_T\!\!\!\!\!\!\!/ \ }

\begin{document}
\def\r#1{\ignorespaces $^{#1}$}
\title{Search for electroweak single top quark production \\ in 
$\mathbf{p\bar{p}}$ collisions at $\mathbf{\sqrt{s}=1.96}$~TeV}
\affiliation{Institute of Physics, Academia Sinica, Taipei, Taiwan 11529, Republic of China}  
\affiliation{Argonne National Laboratory, Argonne, Illinois 60439}  
\affiliation{Institut de Fisica d'Altes Energies, Universitat Autonoma de Barcelona, E-08193, Bellaterra (Barcelona), Spain}  
\affiliation{Istituto Nazionale di Fisica Nucleare, University of Bologna, I-40127 Bologna, Italy}  
\affiliation{Brandeis University, Waltham, Massachusetts 02254}  
\affiliation{University of California at Davis, Davis, California  95616}  
\affiliation{University of California at Los Angeles, Los Angeles, California  90024}  
\affiliation{University of California at San Diego, La Jolla, California  92093}  
\affiliation{University of California at Santa Barbara, Santa Barbara, California  93106}  
\affiliation{Instituto de Fisica de Cantabria, CSIC-University of Cantabria, 39005 Santander, Spain}  
\affiliation{Carnegie Mellon University, Pittsburgh, PA  15213}  
\affiliation{Enrico Fermi Institute, University of Chicago, Chicago, Illinois 60637}  
\affiliation{Joint Institute for Nuclear Research, RU-141980 Dubna, Russia} 
\affiliation{Duke University, Durham, North Carolina  27708}  
\affiliation{Fermi National Accelerator Laboratory, Batavia, Illinois 60510}  
\affiliation{University of Florida, Gainesville, Florida  32611}  
\affiliation{Laboratori Nazionali di Frascati, Istituto Nazionale di Fisica Nucleare, I-00044 Frascati, Italy}  
\affiliation{University of Geneva, CH-1211 Geneva 4, Switzerland}  
\affiliation{Glasgow University, Glasgow G12 8QQ, United Kingdom} 
\affiliation{Harvard University, Cambridge, Massachusetts 02138}  
\affiliation{The Helsinki Group: Helsinki Institute of Physics; and Division of High Energy Physics, Department of Physical Sciences, University of Helsinki, FIN-00044, Helsinki, Finland} 
\affiliation{Hiroshima University, Higashi-Hiroshima 724, Japan}  
\affiliation{University of Illinois, Urbana, Illinois 61801}  
\affiliation{The Johns Hopkins University, Baltimore, Maryland 21218}  
\affiliation{Institut f\"{u}r Experimentelle Kernphysik, Universit\"{a}t Karlsruhe, 76128 Karlsruhe, Germany}  
\affiliation{High Energy Accelerator Research Organization (KEK), Tsukuba, Ibaraki 305, Japan}  
\affiliation{Center for High Energy Physics: Kyungpook National University, Taegu 702-701; Seoul National University, Seoul 151-742; and SungKyunKwan University, Suwon 440-746; Korea}  
\affiliation{Ernest Orlando Lawrence Berkeley National Laboratory, Berkeley, California 94720}  
\affiliation{University of Liverpool, Liverpool L69 7ZE, United Kingdom}  
\affiliation{University College London, London WC1E 6BT, United Kingdom}  
\affiliation{Massachusetts Institute of Technology, Cambridge, Massachusetts  02139}  
\affiliation{Institute of Particle Physics: McGill University, Montr\'{e}al, Canada H3A~2T8; and University of Toronto, Toronto, Canada M5S~1A7}  
\affiliation{University of Michigan, Ann Arbor, Michigan 48109}  
\affiliation{Michigan State University, East Lansing, Michigan  48824}  
\affiliation{Institution for Theoretical and Experimental Physics, ITEP, Moscow 117259, Russia}  
\affiliation{University of New Mexico, Albuquerque, New Mexico 87131}  
\affiliation{Northwestern University, Evanston, Illinois  60208}  
\affiliation{The Ohio State University, Columbus, Ohio  43210}  
\affiliation{Okayama University, Okayama 700-8530, Japan} 
\affiliation{Osaka City University, Osaka 588, Japan}  
\affiliation{University of Oxford, Oxford OX1 3RH, United Kingdom}  
\affiliation{University of Padova, Istituto Nazionale di Fisica Nucleare, Sezione di Padova-Trento, I-35131 Padova, Italy}  
\affiliation{University of Pennsylvania, Philadelphia, Pennsylvania 19104}  
\affiliation{Istituto Nazionale di Fisica Nucleare, University and Scuola Normale Superiore of Pisa, I-56100 Pisa, Italy}  
\affiliation{University of Pittsburgh, Pittsburgh, Pennsylvania 15260}  
\affiliation{Purdue University, West Lafayette, Indiana 47907}  
\affiliation{University of Rochester, Rochester, New York 14627}  
\affiliation{The Rockefeller University, New York, New York 10021}  
\affiliation{Istituto Nazionale di Fisica Nucleare, Sezione di Roma 1, University di Roma ``La Sapienza," I-00185 Roma, Italy} 
\affiliation{Rutgers University, Piscataway, New Jersey 08855}  
\affiliation{Texas A\&M University, College Station, Texas 77843}  
\affiliation{Texas Tech University, Lubbock, Texas 79409}  
\affiliation{Istituto Nazionale di Fisica Nucleare, University of Trieste/\ Udine, Italy}  
\affiliation{University of Tsukuba, Tsukuba, Ibaraki 305, Japan}  
\affiliation{Tufts University, Medford, Massachusetts 02155}  
\affiliation{Waseda University, Tokyo 169, Japan}  
\affiliation{Wayne State University, Detroit, Michigan  48201}  
\affiliation{University of Wisconsin, Madison, Wisconsin 53706}  
\affiliation{Yale University, New Haven, Connecticut 06520}  
\author{D.~Acosta}
\affiliation{University of Florida, Gainesville, Florida  32611} 
\author{J.~Adelman}
\affiliation{Enrico Fermi Institute, University of Chicago, Chicago, Illinois 60637} 
\author{T.~Affolder}
\affiliation{University of California at Santa Barbara, Santa Barbara, California  93106} 
\author{T.~Akimoto}
\affiliation{University of Tsukuba, Tsukuba, Ibaraki 305, Japan} 
\author{M.G.~Albrow}
\affiliation{Fermi National Accelerator Laboratory, Batavia, Illinois 60510} 
\author{D.~Ambrose}
\affiliation{University of Pennsylvania, Philadelphia, Pennsylvania 19104} 
\author{S.~Amerio}
\affiliation{University of Padova, Istituto Nazionale di Fisica Nucleare, Sezione di Padova-Trento, I-35131 Padova, Italy} 
\author{D.~Amidei}
\affiliation{University of Michigan, Ann Arbor, Michigan 48109} 
\author{A.~Anastassov}
\affiliation{Rutgers University, Piscataway, New Jersey 08855} 
\author{K.~Anikeev}
\affiliation{Massachusetts Institute of Technology, Cambridge, Massachusetts  02139} 
% ...
\date{\today}
\collaboration{CDF collaboration}
\noaffiliation
\begin{abstract}
We report on a search for Standard Model $t$-channel and $s$-channel 
single top quark production in $p\bar{p}$ collisions at a center of 
mass energy of 1.96 TeV. We use a data sample corresponding to 
$162\;\mathrm{pb^{-1}}$ recorded by the upgraded Collider Detector at Fermilab. 
We find no significant evidence for electroweak top quark production and
set upper limits at the 95\% confidence level on the production cross section,
 consistent with the Standard Model:
10.1~pb for the $t$-channel, 13.6~pb for the $s$-channel 
and 17.8~pb for the combined cross section of $t$- and $s$-channel. 
\end{abstract}

\maketitle
In $p\bar{p}$ collisions at 1.96 TeV, top quarks are predominantly
produced in pairs via strong interaction processes. 
Within the Standard Model (SM), top quarks are
also expected to be produced singly by the electroweak interaction
involving a $Wtb$ vertex~\cite{stopTheo}.
At the Tevatron, the two relevant production modes are the
$t$- and the $s$-channel exchange of a virtual $W$ boson.
The measurement of the single top cross section is particularly 
interesting because
the production cross section is proportional to $|V_{tb}|^2$,
where $V_{tb}$ is the Cabibbo-Kobayashi-Maskawa (CKM) matrix element
which relates top and bottom quarks.
Assuming three quark generations, the unitarity of the CKM matrix
implies that $V_{tb}$ is close to unity~\cite{PDG2002}.
The most recent next-to-leading order (NLO) calculations, 
assuming $|V_{tb}|=1$,  
predict cross sections of $(1.98\pm0.25)$~pb for the $t$-channel 
and $(0.88\pm 0.11)$~pb for the $s$-channel mode at 
$\sqrt{s}=1.96\;\mathrm{TeV}$~\cite{harris}. 
Using these predictions, a measurement of the 
single top cross section will allow for a direct determination of $|V_{tb}|$.
Single top searches test also exotic models which predict
anomalously altered single top production rates~\cite{anotop}.
Moreover, single top quark processes result in the same final state as
the Standard Model Higgs boson process $WH\rightarrow\ell\nu b\bar{b}$
and therefore impact future searches for the Higgs boson at the 
Tevatron~\cite{higgs}.
In this article we report results of the first search for
single top production in Run 2 at the Tevatron.
Results of searches for single top production at 
$\sqrt{s}=1.8\;\mathrm{TeV}$ (Run 1) can be found in 
Refs.~\cite{stopCDFrun1,stopD0run1}. 

The experimental signature of single top events consists of the $W$ decay
products plus two or three jets, including one $b$ quark jet from the 
decay of the top quark.  
To suppress QCD multijet background
we select only $W\rightarrow \mu\nu_\mu$ and $W\rightarrow e\nu_e$ candidates.
In $s$-channel events we expect a second $b$ quark jet from the $Wtb$ vertex.
In $t$-channel events a second jet originates from the
recoiling light quark and a third jet is produced through the splitting of the
initial-state gluon into a $b\bar{b}$ pair.
Mostly, this third jet escapes detection, since it is produced 
in the high pseudorapidity ($\eta$) regions and 
at low transverse energy ($E_T$)~\cite{eta}. 

This article describes two analyses: (1) a combined search for the
$t$- plus $s$-channel single top signal, (2) a separate search,
where we measure the rates for the two single top
processes individually.
The data sample corresponds to an integrated luminosity of 
$(162\pm10)\,\mathrm{pb^{-1}}$ collected with the upgraded
Collider Detector at Fermilab (CDF II),
which is described elsewhere~\cite{cdfdet}. 
The common event preselection of our two analyses resembles closely the one 
used in the CDF measurement of the $t\bar{t}$ cross section 
reported in Ref.~\cite{ttbarxsec}.
We accept events with evidence for a leptonic $W$ decay: 
(a) missing transverse energy $\EtMiss \,\,> 20\;\mathrm{GeV}$ from the neutrino
and (b) an isolated central electron with $E_T > 20\;\mathrm{GeV}$
or an isolated central muon candidate with $p_T > 20\;\mathrm{GeV}/c$.
An electron or muon candidate is considered isolated 
if the non-lepton $E_T$ in an
$\eta$-$\phi$ cone of radius 0.4 centered around the lepton is less than 10\%
of the lepton $E_T$ or $p_T$.
To remove dilepton events from $t\bar{t}$-production and 
leptonic $Z$ boson decays,
we accept events with only one well identified lepton.
In addition, we veto events if we find a second, loosely identified lepton
candidate that forms an invariant mass with the 
primary lepton between $76\;\mathrm{GeV}/c^2 < M_{\ell\ell} < 106\;\mathrm{GeV}/c^2$.
The jet reconstruction uses a fixed cone of radius 
$\Delta R = \sqrt{\Delta\phi^2 + \Delta\eta^2} = 0.4$.
We count jets  with transverse energy 
$E_T \geq 15\;\mathrm{GeV}$ and $|\eta|\leq 2.8$. 
We only accept $W+2\;$jets events.
%
At least one of these jets must be identified as likely to originate from 
a $b$ quark ($b$-tag) by requiring a displaced secondary vertex within the jet
as measured using silicon tracker information.
The effective coverage of the $b$-tagging ranges up to $|\eta| \lesssim 1.4$.

To optimize our sensitivity, we apply 
a cut on the invariant mass $M_{\ell\nu b}$ of the charged lepton,
the neutrino and the $b$-tagged jet: 
$140\;\mathrm{GeV}/c^2 \leq M_{\ell\nu b} \leq 210\;\mathrm{GeV}/c^2$.
The transverse momentum of the neutrino is set equal to 
the missing transverse energy vector 
$\vec{E\!\!\! / }_T$; 
$p_z(\nu)$ is obtained up to a two-fold ambiguity from the 
constraint $M_{\ell\nu}=M_W$.
From the two solutions we pick the one with lower $|p_z(\nu)|$.
If the $p_z(\nu)$ solution has non-zero imaginary part as a consequence
of resolution effects in measuring jet energies, 
we use only the real part of $p_z(\nu)$.
For the separate search, we subdivide the sample into events with 
exactly one $b$-tagged jet or exactly two $b$-tagged jets. 
For the 1-tag sample, 
we require at least one jet to have $E_T \geq 30\;\mathrm{GeV}$.
We determine the total event detection efficiency $\epsilon_\mathrm{evt}$
for the signal from events generated by the matrix element event generator
MadEvent~\cite{madevent}, followed by parton showering with
PYTHIA~\cite{pythia}
and a full CDF II detector simulation.
MadEvent features the correct spin polarization of the top quark and its
decay products.
For $t$-channel single top production we generated two samples,
one $b+q\rightarrow t+q^\prime$ and one $g+q\rightarrow t+\bar{b}+q^\prime$
which we merged together to reproduce the $p_T$ spectrum of the 
$\bar{b}$ as expected from NLO differential cross section calculations.
This is an improved model compared to the Pythia modelling used in 
the Run I analyses.
The event detection efficiency $\epsilon_\mathrm{evt}$ includes the 
kinematic and 
fiducial acceptance, branching ratios, 
lepton and $b$-jet identification as well as trigger efficiencies.
We combine $\epsilon_\mathrm{evt}$ as given in Table~\ref{tab:acc} with the 
cross 
sections predicted by theory~\cite{harris} and thereby obtain the number of 
expected 
single top events listed in Table~\ref{tab:rates}.
\begin{table}
\caption{\label{tab:acc}Event detection efficiencies in \%.}
\begin{ruledtabular}
  \begin{tabular}{lccc}
  Process   &  Combined &  1-tag & 2-tag \\ \hline
  $t$-channel & 0.89$\pm$0.07 & 0.86$\pm$0.07 & 0.007$\pm$0.002 \\ 
  $s$-channel & 1.06$\pm$0.08 & 0.78$\pm$0.06 & 0.23$\pm$0.02 \\
  \end{tabular}
\end{ruledtabular}
\end{table}

We distinguish between two background components: 
$t\bar{t}$ and non-top background. 
We estimate the $t\bar{t}$ background based on
events generated with PYTHIA, 
normalized to the theoretically predicted cross section of 
$\sigma(t\bar{t})=6.7^{+0.7}_{-0.9}\;\mathrm{pb}$~\cite{ttbar}.
The primary source (62\%) of the non-top background 
is the $W$+heavy flavor processes
$\bar{q}q^\prime\rightarrow Wg$ with $g\rightarrow b\bar{b}$ or 
$g\rightarrow c\bar{c}$, and $gq\rightarrow Wc$.
Additional sources are ``mistags'' (25\%), in which a light-quark 
jet is erroneously identified as heavy flavor, 
``non-$W$''(10\%), e.g. direct $b\bar{b}$ production, and
diboson ($WW$, $WZ$, $ZZ$) production (3\%).
The non-$W$ and mistag fractions are estimated using CDF II data. 
The $W$+heavy flavor rates are extracted from ALPGEN~\cite{alpgen} 
Monte Carlo (MC)
events normalized to data~\cite{ttbarxsec}. 
The diboson rates are estimated from PYTHIA events normalized 
to theory predictions~\cite{dibosonxsec}.
%
The numbers of expected signal and background events are summarized in 
Table~\ref{tab:rates}.
\begin{table}[bt]
\caption{\label{tab:rates}Expected number of signal and background events
  compared with observations.}
\begin{ruledtabular}
  \begin{tabular}{lccc}
    Process    & Combined & 1-tag & 2-tag \\ \hline
    $t$-channel & $2.8\pm0.5$  & $2.7\pm0.4$ & $0.02\pm0.01$ \\
    $s$-channel  & $1.5\pm0.2$  & $1.1\pm0.2$ & $0.32\pm0.05$ \\ \hline
    $t\bar{t}$ & $3.8\pm0.9$  & $3.2\pm0.7$  & $0.60\pm0.14$ \\
    non-top    & $30.0\pm5.8$ & $23.3\pm4.6$ & $2.59\pm0.71$ \\ \hline
    Total Background & $33.8\pm5.9$ & $26.5\pm4.7$ & $3.19\pm0.72$ \\ \hline
    Total Expected     & $38.1\pm5.9$ & $30.3\pm4.7$ & $3.53\pm0.72$ \\  \hline
    Observed   & 42  & 33 & 6 \\ 
  \end{tabular}
\end{ruledtabular}
\end{table}
Having applied all selection cuts we observe 42 events for the combined search,
33 events in the 1-tag sample and 6 events in the 2-tag sample.
Within the uncertainties, the observations are in good agreement with 
predictions.

\begin{figure}[t]
 \begin{flushleft}
  {\footnotesize\sffamily {\bf a)}}
  \end{flushleft}
  \vspace*{-6mm}

  \includegraphics[width=0.4\textwidth]{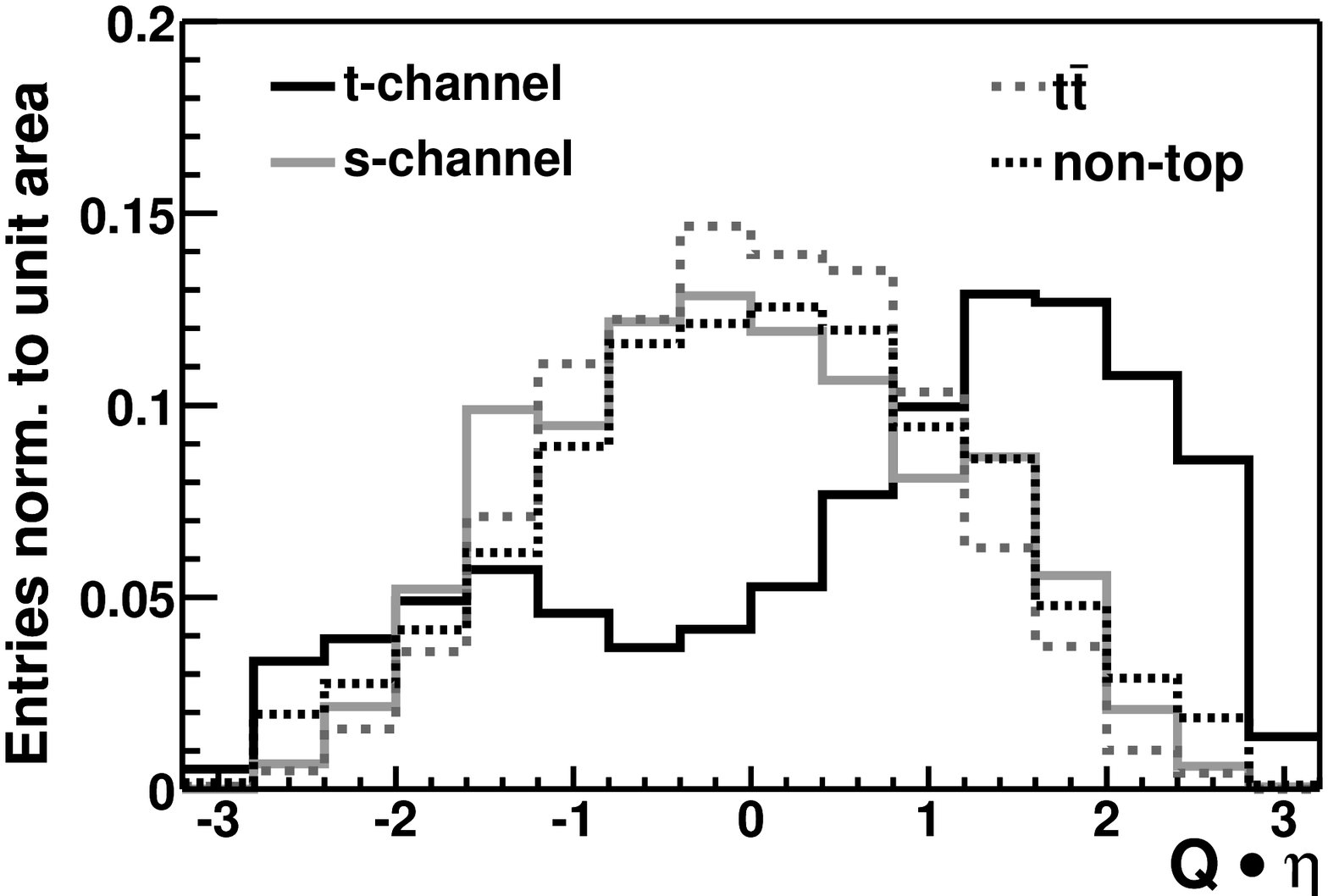}
  \vspace*{-7mm}

  {\flushleft\footnotesize\sffamily {\bf b)}\hspace*{0.47\textwidth} }\\
  \vspace*{-4mm}

  \includegraphics[width=0.4\textwidth]{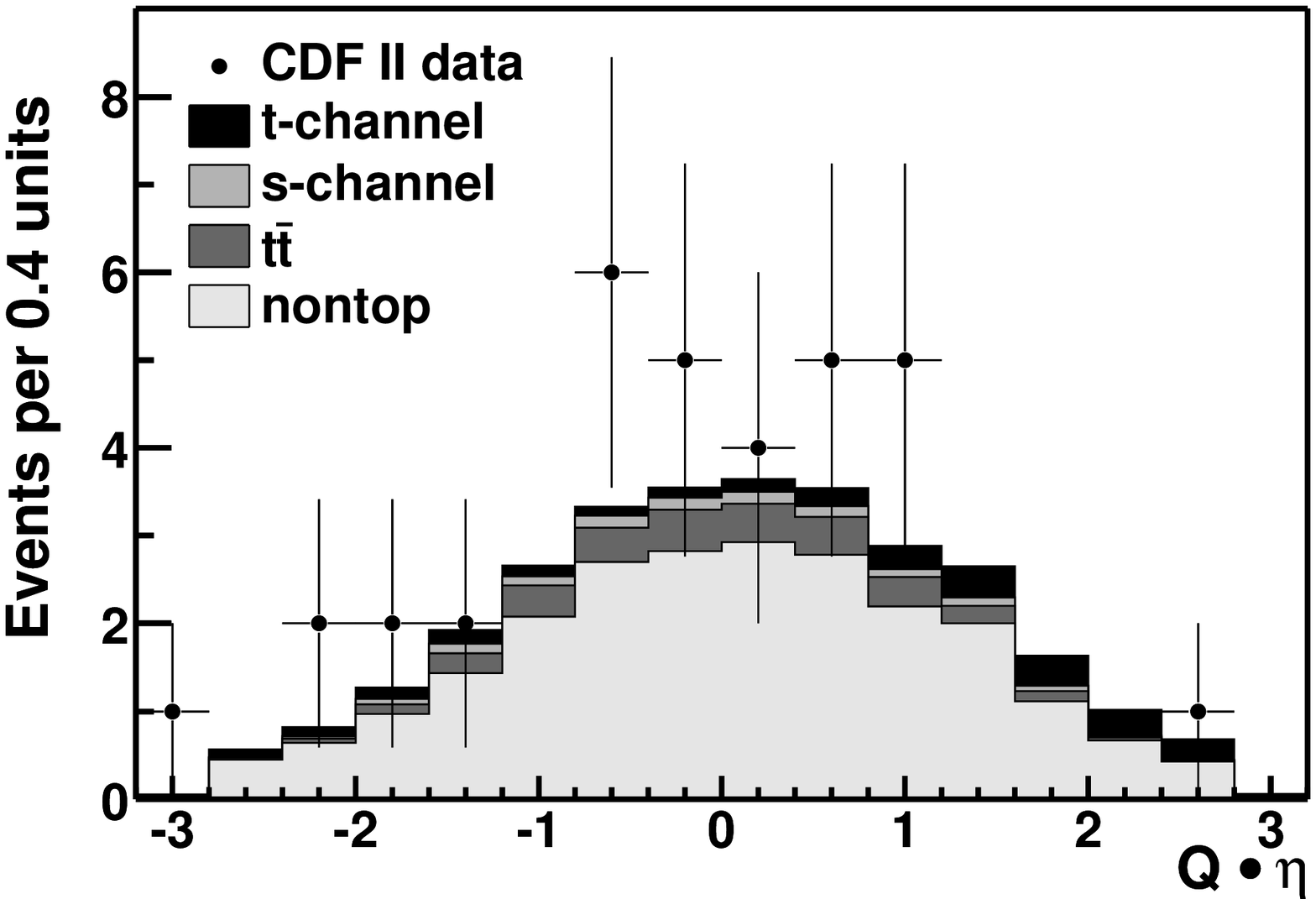}
  \caption{\label{fig:QetaPlots}$Q\cdot\eta$ distributions for a) MC
    templates normalized to unit area, b) Data in the 1-tag sample 
    (33 events) vs. stacked 
    MC templates normalized to the SM prediction.}
\end{figure}

To extract the signal content in data, we use a maximum likelihood technique.
We separate $t$- and $s$-channel events by using the 
$Q\cdot\eta$ distribution which exhibits a distinct asymmetry
for $t$-channel events, see Fig.~\ref{fig:QetaPlots}a. 
$Q$ is the charge of the lepton and $\eta$ is the pseudorapidity 
of the untagged jet.
In Fig.~\ref{fig:QetaPlots}b we show the data versus stacked MC 
templates weighted by the expected number of events in the 1-tag sample.
The separate search defines a joint likelihood function for the
$Q\cdot\eta$ distribution in the 1-tag sample and for the number of events
in the 2-tag sample.
\[ \mathcal{L}_\mathrm{sig} 
   (\sigma_1, ...\,, \sigma_4; \delta_1, ...\,, \delta_7) = 
    \frac{e^{-\mu_{d}} \cdot \mu_{d}^{n_{d}}}{n_{d}!} \cdot 
    \prod_{k=1}^{N_\mathsf{bin}} \frac{e^{-\mu_{k}} \cdot 
    \mu_{k}^{n_{k}}}{n_{k}!} 
 \]

\vspace*{-4mm}

\[  \cdot \prod_{\stackrel{j=1}{j\neq sig}}^{4} 
    G (\sigma_{j}; \sigma_\mathrm{SM,j}, 
    \Delta_{j})
    \cdot \prod_{i=1}^{7} G(\delta_{i}; 0, 1) \]
Four processes are considered and labeled by the index $j$:
$t$-channel ($j=1$), $s$-channel ($j=2$), $t\bar{t}$ ($j=3$), and 
non-top ($j=4$).
The corresponding cross sections are denoted $\sigma_j$.
The background cross sections 
are constrained to their SM prediction $\sigma_\mathrm{SM,j}$ 
by Gaussian priors 
of width $\Delta_3=23\%\,\sigma_\mathrm{SM,3}$ for $t\bar{t}$ 
and $\Delta_4=20\%\,\sigma_\mathrm{SM,4}$ for non-top. 
The index ``sig'' denotes the signal process,
which is s- or $t$-channel, respectively.
The $\mu_k$ are the mean number of events in bin $k$ of the 
$Q\cdot\eta$ histogram ($N_{bin} \equiv$ number of bins), 
while $\mu_d$ is the mean number of events in the 2-tag sample. 
$n_k$ and $n_d$ are the event numbers observed in data, respectively.
Seven sources of systematic uncertainties are considered
in the likelihood function:
(1) jet energy scale (JES), (2) initial state
radiation (ISR), (3) final state radiation (FSR), 
(4) parton distribution functions (PDF),
(5) the choice of signal MC generator, (6) the top quark mass, (7)
trigger, identification and $b$-tagging efficiencies and the luminosity.
The relative strength of a systematic effect due to source $i$ is parameterized
by the variable $\delta_i$.
Systematic effects change the acceptance and influence 
the shape of the $Q\cdot\eta$ distribution.
When calculating $\mu_{k/d}$ we take the systematic shifts in the
acceptance and in the shape of the template histograms,
and their full correlation into account.
All variables except the signal cross section
$\sigma_{sig}$ are constrained
to their expected values by Gaussian functions $G(x; x_0, \Delta_x)$
of mean $x_{0}$ and width $\Delta_{x}$.
The largest uncertainties are on the $b$-tagging efficiency (7\%),
luminosity (6\%), top quark mass (4\%) and JES (4\%).
The effect of uncertainty in the JES is evaluated by 
applying energy corrections
that describe $\pm1\,\sigma$ variations.
Systematic uncertainties due to the modeling of ISR and FSR are obtained from 
MC samples that describe variations in these effects.
To evaluate the uncertainty associated with the choice of a specific 
parametrization of PDF we investigated several
PDF sets and took the maximum deviation (MRST72) from our standard 
PDF set (CTEQ5L). We estimate the uncertainty associated with the choice 
of single top MC generator using samples generated with 
TopReX~\cite{toprex}.
The values of acceptance uncertainties for the single top processes 
are summarized in Table~\ref{tab:systuncert}.
\begin{table}
\caption{\label{tab:systuncert}
 Fractional changes in $\epsilon_\mathrm{evt}$
 of single top processes in \%. $\epsilon_\mathrm{trig}$ is the trigger
 efficiency, $\epsilon_\mathrm{ID}$ the lepton identification efficiency.}
\begin{ruledtabular}
  \begin{tabular}{ccccc}
   $i$& Source     & $t$-channel & $s$-channel & Combined \\ \hline 
    1 & JES              & $^{+2.4}_{-6.7}$ & $^{+0.4}_{-3.1}$ & 
                           $^{+0.1}_{-4.3}$ \\ 
    2 & ISR              & $\pm1.0$ & $\pm0.6$ & $\pm1.0$ \\ 
    3 & FSR              & $\pm2.2$ & $\pm5.3$ & $\pm2.6$ \\ 
    4 & PDF              & $\pm$4.4 & $\pm$2.5 & $\pm$3.8 \\ 
    5 & Generator        & $\pm$5 & $\pm$2 & $\pm$3 \\ 
    6 & Top quark mass   & $^{+0.7}_{-6.9}$ & -2.3 & -4.4 \\ 
    7 & $\epsilon_\mathrm{trig}$, $\epsilon_\mathrm{ID}$, luminosity
                         & $\pm9.8$ & $\pm9.8$ & $\pm9.8$ \\
  \end{tabular}
\end{ruledtabular}
\end{table}

To measure the combined $t$-channel plus $s$-channel signal in data,
we use a kinematic variable whose distribution is very similar for the
two single top processes, but is different for background processes:
$H_T$, which is the scalar sum  of $\EtMiss\;$ and the transverse 
energies of the lepton and all jets in the event.
\begin{figure}[tb]
\begin{center}
  \includegraphics[width=0.4\textwidth]{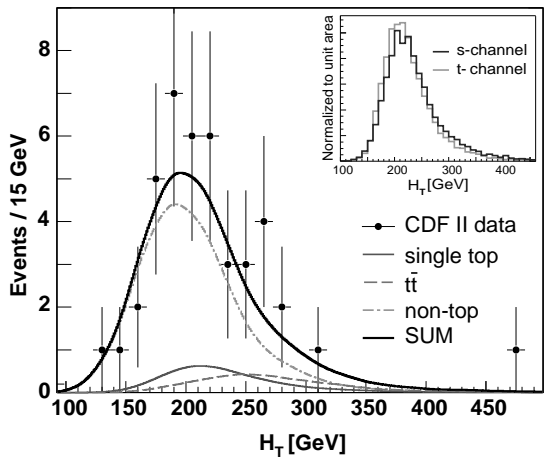}
  \caption{\label{fig:HTtemplates} $H_T$ distribution for data (42 events) 
   in the combined search compared with smoothed MC predictions for signal 
   and background.}
\end{center}
\end{figure}
We use a likelihood function similar to that in the separate search. 
One difference is that in the combined search the $H_T$ distributions are 
smoothed.
In Fig.~\ref{fig:HTtemplates} we show the $H_T$ distribution observed 
in data compared to the SM prediction.

We perform Monte Carlo experiments to estimate our {\it a priori} sensitivity
assuming the SM signal cross sections. 
For each experiment we integrate out all nuisance parameters 
({\it i.e.}, all variables except $\sigma_{sig}$) from the full likelihood 
function and thereby construct the marginalized likelihood
$\mathcal{L}^*_{sig}(\sigma_{sig})$.
$\mathcal{L}^*_{sig}$ is normalized and interpreted as 
posterior probability
density function $p(\sigma_{sig})$.
We calculate the upper limit at the 95\% C.L. using a Bayesian
method assuming a prior probability density,
which is 0 if $\sigma_{sig} < 0$ and 1 if $\sigma_{sig} \geq 0$.
The median of the expected upper limits defines our sensitivity
and is stated in Table~\ref{sumar}.
\begin{table}[!t]
\caption{\label{sumar}
Upper limits at the 95\% C.L. and most probable values (MPV) of 
single top cross sections in pb.} 
\begin{ruledtabular}
\begin{tabular}{lccc}
  & $t$-channel & $s$-channel & Combined \\ \hline
expected limit & 11.2 & 12.1 & 13.6 \\
observed limit & 10.1 & 13.6 & 17.8 \\
MPV $\pm$ HPD & $0.0^{+4.7}_{-0.0}$ & $4.6^{+3.8}_{-3.8}$ &
           $7.7^{+5.1}_{-4.9}$ \\
\end{tabular}
\end{ruledtabular}
\end{table}
We calculate 
the posterior probability densities for CDF II data 
and obtain
the most probable values (MPV) and highest posterior density
(HPD) intervals~\cite{PDG2002} as given
in Table~\ref{sumar}. Within the statistical uncertainty these
results are compatible with SM predictions.
We find upper limits of 10.1~pb at the 95\% C.L.
for the $t$-channel cross section
and 13.6~pb for the $s$-channel.
For the combined search 
we find an upper limit of 17.8~pb at the 95\% C.L..

In summary, we find no significant evidence for electroweak 
single top quark production
in $(162\pm10)\;\mathrm{pb^{-1}}$ of integrated luminosity recorded with CDF II.
We set the first limits on single top cross sections in $p\bar{p}$ collisions
at $\sqrt{s}=1.96\;\mathrm{TeV}$ in Run 2 at the Tevatron.
If compared with Run 1 results the upper limits on $t$-channel and $s$-channel
single top quark production are considerably improved by
28\% ($t$-channel) or 20\% ($s$-channel), respectively.
We have introduced improved Monte Carlo modeling for single top
and a fully Bayesian treatment of systematic uncertainties in the likelihood 
function which are important steps for future
analyses aiming for the observation of single top quark production. 

\begin{acknowledgments}
We wish to thank the Fermilab staff and the technical staffs of the
participating institutions for their vital contributions.
This work was supported by the U.S. Department of Energy and National
Science Foundation; the Italian Instituto Nazionale di Fisica Nucleare;
the Ministry of Education, Culture, Sports, Science and Technology of Japan;
the Natural Sciences and Engineering Research Council of Canada; 
the National Science Council of the Republic of China; 
the Swiss National Science Foundation;
the A.P. Sloan Foundation; 
the Bundesministerium f\"ur Bildung und Forschung, Germany; 
the Korean Science and Engineering Foundation and the Korean Research Foundation;
the Particle Physics and Astronomy Research Council and the Royal Society, UK;
the Russian Foundation of Basic Research; 
the Comisi\'on Interministerial de Ciencia y Tecnolog\'{\i}a, Spain;
and in part by the European Community's Human Potential Program under contract
HPRN-CT-20002, Probe for New Physics. 
We acknowledge the help of T. Stelzer
and S. Slabospitsky for the generation of MadEvent and TopReX MC samples.
We thank S. Mrenna and Z. Sullivan for useful discussions.
\end{acknowledgments}


\begin{thebibliography}{}
\bibitem{eta}The polar angle $\theta$ is measured 
w.r.t. the proton beam direction. The pseudorapidity 
is defined as $\eta\equiv-\ln\tan(\theta/2)$. $E_T = E\,\sin(\theta)$.
\end{thebibliography}
\end{document}